\documentclass[twocolumn,showpacs,amsmath,amssymb,prb,floatfix]{revtex4}

\usepackage{graphicx}
\usepackage{dcolumn}

\begin{document}


\title{Power laws in a 2-leg ladder of interacting spinless fermions.}

\author{L.~G. Caron}
\email{caron@physique.usherb.ca}
\author{C. Bourbonnais}
\email{cbourbon@physique.usherb.ca}
\affiliation{D\'{e}partement de physique and Centre de recherche
sur les propri\'{e}t\'{e}s \'{e}lectroniques de mat\'{e}riaux
avanc\'{e}s (CERPEMA),\\
Universit\'{e} de Sherbrooke, Sherbrooke, QC J1K 2R1, Canada.}

\date{\today}

\begin{abstract}
We use the Density-Matrix Renormalization Group to study the
single-particle and two-particle correlation functions of spinless
fermions in the ground state of a quarter-filled ladder. This
ladder consists of two chains having an in-chain extended Coulomb
interaction reaching to third neighbor and coupled by inter-chain
hopping. Within our short numerical coherence lengths, typically
reaching ten to twenty sites, we find a strong renormalization of
the interchain hopping and the existence of a dimensional
crossover at smaller interactions. We also find power exponents
for single-particle hopping and interchain polarization consistent
with the single chain. The total charge correlation function has a
larger power exponent and shows signs of a crossover from incoherent
fermion hopping to coherent particle-hole pair motion between chains.
There are no significant excitation energies.
\end{abstract}


\maketitle


\section{Introduction}

The theory of quasi-one-dimensional conductors\cite
{Solyom79,Emery79,Firsov85,Brazovskii85b,Bourbon91,Bourbon95,Schulz95,%
Schulz00,Voit95}
has shown that there are dimension-specific aspects not observed
in conventional three-dimensional solids. Aside from the
interaction dependent power law behavior of single-particle and
pair response functions and the well documented spin-charge
separation, there is the renormalization of the transverse hopping whose
impact on the description of real materials is much debated\cite{Bourbon99,%
Biermann01}. As discussed in
length in Ref. \onlinecite{Bourbon91}, strong Coulomb interactions can
dramatically reduce the effective value of the transverse hopping
and retard the dimensionality crossover from a one-dimensional
(1D) to a two- or three-dimensional conductor. The simplest
theoretical testing ground for this idea is a 2-leg ladder
consisting of interacting spinless fermions on two chains coupled
by a transverse hopping $t_{\bot }$. It is in principle possible
to study the putative renormalization of $t_{\bot }$. This has
been done using various approaches, among which exact
diagonalization\cite{Capponi98}, momentum-space
renormalization\cite {Fabrizio93}, and
bosonization\cite{Suzumura95,Nersesyan93,Yakovenko92,Donohue01}. In all
these papers, renormalization of the interchain hopping is
confirmed. What we propose is a numerical calculation of this
two-chain problem at quarter-filling using the efficient density-matrix
renormalization group\cite {Noack92,White92,White93,DMRGreview99}
(DMRG) in order to directly measure power law exponents and the
effective value of the interchain hopping. Although the DMRG has
recently been tried\cite{Yonemitsu01} on spinless fermions, the analysis
focused on a half-filled ladder and the nearest neighbor current correlations.

We shall first present the model Hamiltonian we shall be using
throughout and, second, the proposed DMRG procedure. Thirdly, we
shall validate our approach on the single chain situation. We
thereafter present the results for two chains and discuss the
results in the light of the various theoretical treatments. A
brief summary follows.

\section{Hamiltonian}

We shall use the model Hamiltonian proposed by Capponi {\it et
al.}\cite{Capponi98} for two quarter-filled chains of spinless
fermions that interact within each chain through a finite extent
Coulomb potential and can hop between chains through the hopping
term $t_{\bot }$. \ The Hamiltonian is
\begin{eqnarray}
H &=&-\sum_{j,\beta }\left( c_{j+1,\beta }^{\dagger }c_{j,\beta
}+h.c.\right) +\sum_{j,\beta ,r}V(r)\,n_{j+r,\beta }\,n_{j,\beta
}\allowbreak \mathstrut  \nonumber \\
&&-t_{\bot }\sum_{j}\left( c_{j,1}^{\dagger }c_{j,2}+h.c.\right)
\label{eq:H}
\end{eqnarray}
\newline
where $c_{j,\beta }$ annihilates a fermion at site $j$
($j=1,...,N$) on chain $\beta $ ($\beta =1,2$), $n_{j,\beta }$ is
the occupancy at the same site, and $V(r)=2V/(r+1)$ is the
intra-chain interaction between first, second, and third
neighboring sites ($r=1,2,3)$ with $V$ as the interaction
strength. We have set the intra-chain hopping element equal to
one.

We have chosen an interaction to third nearest neighbor because
the work of Capponi showed that the single-fermion exponent
$\alpha $, characterizing the long-range single-chain inter-site
transfer function
\begin{equation}
C1(j,r)=\langle c_{j+r}^{\dagger }c_{j}\rangle \varpropto
r^{-(1+\alpha )}, \label{eq:C1}
\end{equation}
can become very large ($\alpha \lesssim 1.5$ for $V\leq 6$). This
power exponent is responsible for the perhaps better known
singularity in the momentum distribution at the Fermi level of
Luttinger liquids. In the limit of small $\alpha $, one has
$\left[ n(k)-n(k_{F})\right] \sim \left| k-k_{F}\right| ^{\alpha
}\mathrm{sign}(k_{F}-k)$. Large values of $\alpha $ will be easily observed
and are expected to lead to much more important effects on the
effective value of $t_{\perp }$. Large values of $\alpha $ are
also synonymous with strong variations in the stiffness $K$. The
two are related through the relation
\begin{equation}
\alpha = \frac{1}{2} \left( K+1/K-2\right)  \label{eq:alpha}
\end{equation}
for spinless fermions on a chain. Consequently, the power law
exponents of the various response functions, which are related to
$K$, will also be strongly affected.

\section{Density-Matrix Renormalization Group}

The exact diagonalization of Eq.(\ref{eq:H}) by
Capponi\cite{Capponi98} was for short chains of up to twenty
sites. Needless to say that some sort of extrapolation procedure,
finite-size scaling in this case, was needed to obtain ground
state information in the thermodynamic limit. We have chosen to
use the DMRG since much longer chains can be studied. This, in
principle at least, should take the system much closer to the
thermodynamic limit and do away with the requirement of performing
a finite-size scaling analysis.

Another shortcoming of short chain lengths has to do with a
``dimensionality'' crossover in the interchain
hopping\cite{Bourbon91,Suzumura95,Nersesyan93}. For
temperatures or frequencies larger than approximately $\left|
t_{\perp }\right| $, the chains do not ``see'' the interchain
hopping, which is incoherent or diffusive, and they are
approximately independent. But in the opposite situation, the
chains are tightly coupled and they form bands having transverse
dispersion. Let us illustrate this in the situation of quarter
filling for $V=0$ and an even number of fermions. An exact
solution to two coupled chains is available. The states are
labelled by $k_{m}=\pi m/(N+1)$ where $1\leq m\leq N$ and have
energy $E_{\pm }(k_{m})=-2\cos (k_{m})\pm \left| t_{\perp }\right|
$. For $t_{\perp }=0$, all levels up to $m=m_{F}=N/4$ are filled
with $N/2$ fermions. There is no interchain hopping. As $\left|
t_{\perp }\right| $ increases, this remains so until
$E_{+}(k_{m_{F}})=E_{-}(k_{m_{F}+1})$, that is until $\left|
t_{\perp }\right| \approx (\pi v_{F}/2)/(N+1)$. Here $v_{F}$ is
the Fermi velocity equal to $\sqrt{2}$ in our units. At this point
there is a sudden change in interchain hopping since the two top
levels below the
Fermi level are $E_{-}$ states. The total interchain hopping energy $%
E_{\perp }$ is now $-2\left| t_{\perp }\right| .$ The next jump occurs at $%
\left| t_{\perp }\right| \approx 3(\pi v_{F}/2)/(N+1)$ when $%
E_{+}(k_{m_{F}-1})=E_{-}(k_{m_{F}+2})$, after which $E_{\perp
}=-4\left| t_{\perp }\right| $. At a given $\left| t_{\perp
}\right| ,$ the jumps occur at $N\approx \left[ (\pi
v_{F}/2t_{\perp })(2p-1)-1\right] $ for $p=1,2...$, when $E_{\perp
}=-2p\left| t_{\perp }\right| .$ Taking $\left| t_{\perp }\right|
=0.1$ for example, $(\pi v_{F}/2t_{\perp })=10\pi /\sqrt{2}\sim
22$. This is a large value. It is therefore difficult to attempt
finite-size scaling or the DMRG under such conditions. One can
only hope of reaching the thermodynamic limit for $N\gg (\pi
v_{F}/2t_{\perp })$. This behavior is surely attenuated in the
presence of the Coulomb interaction which scrambles the spectrum.
But short chains remain unpredictable because of the discrete
energy spectrum. Thus the longer chain lengths obtainable with the
DMRG would circumvent this potential numerical distortion. In the
event that $t_{\perp }$ renormalizes to much smaller values than
the bare one, this crossover phenomenon might even prove
cumbersome to the DMRG. In order to avoid potential problem we
chose to use the finite system algorithm proposed by
White\cite{White92,White93,DMRGreview99}, projecting out the ground
state of the superblock. At a given $V$, we
started with the procedure with largest value of $\left| t_{\perp
}\right| $ we wished to consider, $0.5$ in all cases, and then
gradually decreased its value using the previous solution as a
seed. For each set of parameters, the iterations stopped when the
discontinuity in the ground state energy and the excitation energy
$E_{x}$ at mid-course, when all block information has just been
refreshed, were judged acceptable. This was typically for 3
iterations. We used open boundary conditions since periodic
boundary conditions lead to unacceptably large truncation errors.

Let us lastly comment on the number of central sites to use in the
DMRG algorithm. The long-range character of the Coulomb
interaction complicates the calculations. For two inner
double-sites (two sites on each chain), the computation resources
(execution time and memory requirements) scale roughly as
$4^{2}\left( N_{B}\right) ^{4}$ where $N_{B}$ is the number of
sites in each of the side blocks. This comes from the coupling of
the two blocks through $V(r=3)$. If one instead chooses to have
three inner
double-sites, the blocks no longer couple and the resources scale as $%
4^{3}48\left( N_{B}\right) ^{3}$ which comes from the coupling of
the inner
sites to each block. We have found that the resources are similar for $N_{B}%
\sim 100$ in qualitative agreement with this crude analysis. We
have
used two inner sites for the values of $N_{B}=42,64,96$ and three for $%
N_{B}=128$.

\section{Single Chains}

It is of utmost importance to test our DMRG procedure on simpler
single-chain problems. There are two delicate aspects that need to be
validated, both linked to the open-ended boundary conditions. The
first one has to do with the value of $N$ that can be chosen for a
specific band filling. The second one concerns the numerical
treatment that must be done on the data in order to generate
information for infinite-length chains.

\subsection{\label{sec:clbf}Chain length and band filling}

\begin{figure}
\includegraphics[trim=0in 2.2in 0.5in 2.9in, scale=.42,clip]{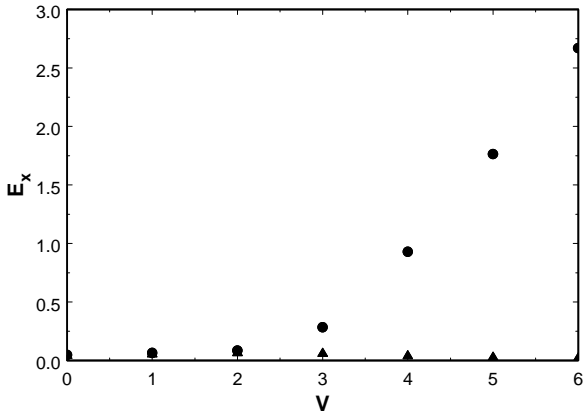}
\caption{\label{fig1}Ground state excitation energy of a
half-filled chain of spinless fermions with nearest-neighbor
interaction strength $V$. The number of sites is 151 (circles) and
152 (triangles).}
\end{figure}

The sensitivity to open ended boundary conditions can be
illustrated for a chain of spinless fermions with an interaction
extending only to nearest neighbor sites ($V(r)=V\delta _{r,1}$)
near half filling. The ground state
and the excitation energy are completely different for the $N=2N_{f}$ and $%
N=2N_{f}-1$ situations, where $N_{f}$ is the number of fermions.
Fig. \ref {fig1} shows the ground state excitation energy in both
situations for a calculation with $N_{B}=42$ block states and
$N=151,152$. Although $N_{B}$ seems small, the truncation error
was nevertheless smaller than $3\times
10^{-7}$ for an open ended chain. A gap develops for the case $N=151$ and $%
N_{f}=76$ but not for the other. As for the ground states, they
show a site occupancy $n_{j}=\overline{n}+A_{m}\cos (\pi j-\theta
_{j})$ that is alternating between a large and a small value. This
basic pattern is to be expected for a broken symmetry state with a
repulsive interaction. But while
the modulation phase $\theta _{j}$ is a constant for $N=151$, it varies for $%
N=152$. We find $\theta \sim \pi j/N$ in a calculation where the
interaction is introduced right away in the first iteration of the
finite size algorithm (sudden turn on) but $\theta \sim 3\pi j/N$
when a first set of iterations is done with $V=0$ and then $V$ is
introduced in the second set of iterations, using the first one as
seed (gradual turn on). The ground state energy is lowest in the
latter situation. The alternating occupancy in the ground state
can be understood by looking at the large $V$
limit when one might expect the fermions to segregate on alternating sites $%
\left| ...1010...\right\rangle $. It is the boundary conditions
that will determine the modulation phase. For $N=151$, one would
expect $\left| 101...101\right\rangle $ to be the stablest
situation, with $\theta _{j}=0$,
and this would explain a uniform modulation and the excitation gap. But for $%
N=152$, the chain will spontaneously create a kink soliton (phase shift of $%
\pi $ from end to end) $\left| 101...010\right\rangle
\Longrightarrow \left| 101...001\right\rangle $ which can then
redistribute itself and lead to a gapless excitation situation.
This is found for the sudden turn on. However, it is also possible
for the chain to generate additional kink-antikink excitations.
The results for the gradual turn on confirm this and show that
this situation is more stable. Just how many kink-antikink pairs
would be generated is impossible to figure out with the DMRG.
Curiously, the DMRG seems unable to yield an unambiguous ground
state when $N=2N_{f}$ for open boundary conditions.

The single-chain Hamiltonian we have just been studying is akin to
the XXZ problem for a spin 1/2 chain\cite{Haldane80}. This spin
Hamiltonian can be transformed, using the
Wigner-Jordan transformation, into our problem with
$J_{z}=V$ and $J_{x}=J_{y}=2$ except for end terms $\frac{1}{2}V(n_{1}+n_{N})$,
involving the first and last sites, that occur for open
boundary conditions. The XXZ chain is known to develop a gap for $%
J_{z}>J_{x} $, that is for $V>2$. For $N=2N_{f}$, the ground state
is degenerate, $\left| 101...010\right\rangle $ and $\left|
010...101\right\rangle ,$ has no soliton because of the end site
repulsion and has a gap. It is the pressure applied to the
fermions by the end terms that insures the presence of an
excitation energy and a uniform modulation amplitude in this
situation. In the case of our ladder with $N=2N_{f}-1$, it is the
shorter length that adds extra pressure to the fermions and
similarly leads to a gapped situation.

\begin{figure}
\includegraphics[trim=0in 2.2in 0.5in 2.9in, scale=.42,clip]{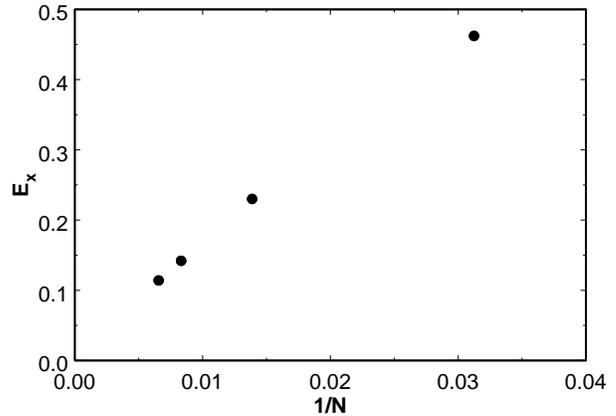}
\caption{\label{fig2}Ground state excitation energy of a
quarter-filled chain of spinless fermions with third-neighbor
interaction strength $V=6$ as a function of the reciprocal of the
chain length.}
\end{figure}

In view of this dichotomy with respect to occupation near half
filling, it is legitimate to ask if this sensitivity persists near
quarter filling. To
this end, we did a limited incursion with {\it gradual turn on} at $V=6$, $%
N_{f}=38$, $N=149,150,152$, and for first neighbor ($r=1$), second neighbor (%
$r=1,2$), and third neighbor ($r=1,2,3$) interactions. The ground
state
energy per fermion for a given interaction range decreases slightly with $N$%
. The effective constraining ``pressure'' when $N=4N_{f}-m$
($m=1,2,3)$ can explain this. The ground state excitation energy
$E_{x}$ remains small going from 0.046 to 0.11 as the range
increases and is insensitive to $N$. Judging from Fig. \ref{fig1},
this is not a significant gap and is due to the finite length of
the chains. Fig. \ref{fig2} shows the excitation energy as a
function of the inverse of the chain length for a third-neighbor
interaction strength $V=6$ and $N_{B}=42$. The extrapolated gap
for $N\rightarrow \infty $ is indeed negligibly small. There is,
however, a variation of the modulation of the site occupancy of
the form $n_{j}=\overline{n}+A_{m}\cos (\pi j/2-\theta _{j}).$
Indeed, for $N=152$, we find $\theta _{j}=(3\pi j)/(2N)$. This
situation is the generalization of the one seen above for half
filling and gradual turn on condition, the quantum of phase shift
being $\pi /2$ instead of $\pi $. What this $\theta _{j}$ means is
that the Fermi momentum is downward shifted from its exact
quarter-filled value $k_{F}=(\pi /4)(1-3/N)$ due to fermions being
pushed to the ends. The question spontaneously arises as to any
possible detrimental effect of such modulation on the correlation
functions. We shall answer this in the following subsection.

\subsection{\label{sec:dataproc}Data processing}

We have just found that a modulation in the site occupancy results
from the open boundary conditions which pin fermions at the end
site and lead to a broken symmetry state. This modulation
obviously makes it non trivial to get occupancies (the
$\overline{n}$) or correlation functions resembling those for
periodic boundary conditions or an infinite chain. Taking the
occupancy as an example, for the quarter-filled situation, it can
be seen that the modulated part $A_{m}\cos (\pi j/2-\theta _{j})$
can be made less annoying by averaging over the natural four site
cycle. Indeed, if we define for instance $\overline{n}_{j}=
\frac{1}{4}\sum_{i=0}^{3}n_{j+i}=\overline{n}-A_{m}\left[ \cos
\left( \pi j/2-\theta _{j}\right) +\sin (\pi j/2-\theta
_{j})\right] (3\pi /4N)$, one immediately sees that the modulation
is reduced by a factor of order $N^{-1}$. A second averaging,
$\bar{n}_{j}=\left( 1/16\right)
\sum_{i=0}^{3}\sum_{m=0}^{3}n_{j+i+m}$, would further reduce it by another $%
N^{-1}$. One can thus, for all practical purposes, remove the
effect of the modulation for long chains, thus our choice of long
chain lengths. We have used throughout what follows double
averaging with interactions reaching to third neighbor at quarter
filling.

\begin{figure}
\includegraphics[trim=0in 2.2in 0.5in 2.9in, scale=.42,clip]{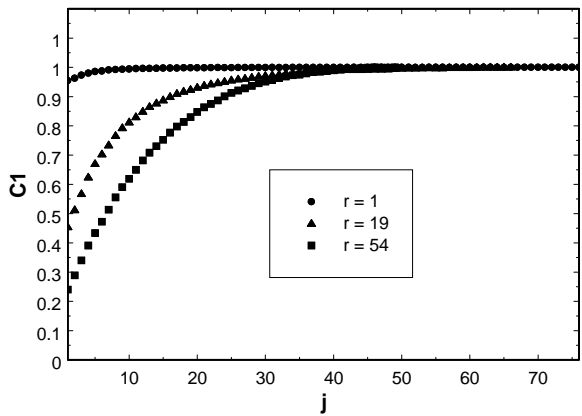}
\caption{\label{fig3}Doubly averaged single-fermion transfer
function for a quarter-filled chain normalized to its value at the
maximum as a function of
the site of origin $j$. The plots stop at $j=(N-r)/2.$ We have chosen $N=152$%
, $N_{B}=42$, $V=6$, and three different transfer distances (see
legend).}
\end{figure}

But, unfortunately, the open boundary conditions produce yet
another deformation. The smoothed quantities, like $\overline{n}$,
are not global quantities but rather local ones. They vary along
the chain, the more so the closer a site is to the ends. We can
illustrate this by looking at the profile of the single-fermion
transfer function $\overline{C1(j,r)}$ defined in Eq.
(\ref{eq:C1}) using double averaging, $N=152$, $V=6$ and
$N_{B}=42$. The plots are for $1\leq j\leq (N-r)/2$ since the
results are symmetrical about this last value. Fig. \ref{fig3}
shows this function for $r=1,19,54$, normalized to
$\overline{C1((N-r)/2,r)}$. The healing distance increases with
$r$ and is at the scale of $r$. It is quite obvious that the ends
can have dramatic effects at the larger values of $r$. We have
used this optimal positioning at $j=(N-r)/2$ in all calculations
reported below of quantities dependent on $r$.

But are averaging and optimal positioning sufficient to obtain
results corresponding to the thermodynamic limit? We can check
this by extracting
the power law exponent of the inter-site transfer function $C1(r)=\overline{%
C1((N-r)/2,r)}$. In order to do so, one needs to have an analytic
function with which to fit the data for $C1(r)$. From the known
form of the single-particle Green's function of a Luttinger
liquid\cite {Schulz00,Voit95}, this function would read as
\begin{equation}
C1(r)=\left( \frac{\overline{n}\sin (k_{F}\,r)}{k_{F}\,r}\right)
\left( \frac{1}{1+(r/\Lambda )^{2}}\right) ^{\alpha /2}
\label{eq:C1a}
\end{equation}
where $k_{F}$ is the Fermi momentum. This form was obtained from
bosonization and $\Lambda $ is a characteristic cutoff parameter.
This form, however, has to be modified to account for the DMRG
procedure on a finite lattice. We find that the gap associated
with a finite $N$ introduces a
numerical coherence length $\xi $ in much the same way that a temperature $%
T\varpropto N^{-1}$ would. We thus propose the substitution\cite{Emery79,Frahm%
90} $r\rightarrow \xi \sinh (r/\xi )$ and thus the following form
\begin{equation}
C1(r)=\left( \frac{\overline{n}\sin (k_{F}\,r)}{k_{F}\,\xi \sinh (r/\xi )}%
\right) \left( \frac{1}{1+(\xi \sinh (r/\xi )/\Lambda
)^{2}}\right) ^{\alpha /2}.  \label{eq:C1b}
\end{equation}
We calculated $C1(r)$ for a chain having $N=152$, $V=1,2...6$, and $N_{B}=42$%
. The truncation error was less than $10^{-8}$. The fitted
parameters, in
the range $1\leq r\leq 100$, are in Table \ref{table1}. It is seen that $%
k_{F}\sim (\pi /4)(1-3/N)$ as expected. The power law exponents $%
\alpha $ are those found by Capponi \textit{et
al}.\cite{Capponi98} although they seem systematically larger by
5-8 \%. The coherence lengths are satisfactorily quite large,
giving added credibility to our fitting function. The Pearson
correlation coefficient\footnote{%
The Pearson correlation coefficient is equal to the covariance of
the two
variables divided by the products of their standard deviations.} between $%
\xi $ and $E_{x}^{-1}$ is 0.986 indicating that the two parameters
are strongly correlated. Finally, the cutoff $\Lambda$ is of order
one, the lattice parameter, as one would expect for fermions on a
lattice. It can thus be concluded that the fitting function, Eq.
(\ref{eq:C1b}), is quite satisfactory, with an error margin of
order 5 \%, considering the large number of adjustable parameters.
The thermodynamic limit is thus recovered albeit slightly
handicapped by a numerical coherence length $\xi $. Now how does
the $2k_{F}$ charge density fluctuation response function turn
out? We also calculated the correlation function
$C2(r)=\overline{C2((N-r)/2,r)}$ where
\begin{equation}
C2(j,r)=\left\langle \left( n_{j+r}-\left\langle
n_{j+r}\right\rangle \right) \left( n_{j}-\left\langle
n_{j}\right\rangle \right) \right\rangle \text{ \ .}
\label{eq:C2}
\end{equation}
This function measures the correlation between occupancy
(proportional to charge) fluctuations on site $j$ and $(j+r)$. It
has the advantage of getting at the true fluctuation correlations
in a broken symmetry state. There are, however, many wavenumber
contributions to two-fermion Green's
functions\cite{Schulz00,Voit95}. One expects $q=0$ and $4k_{F}$
contributions aside from the sought $2k_{F}$ correlations. We observe that $%
C2(r)$ has a fast oscillating part $B_{f}(r)$ and a slow
modulation amplitude $A_{s}(r)$, such that $C2(r)\approx
A_{s}(r)B_{f}(r)$. What we did was to exponentiate the data
$C2(r)\exp \left( -\ln \left( \left| A_{s}(r)\right| \right)
\right) $, do a fast Fourier transform, remove the unwanted
contributions, and unexponentiate back the remaining $2k_{F}$
contribution. For spinless fermions, the charge correlation function at $%
2k_{F}$ behaves like $r^{-2K}$ where $K$ is the stiffness defined
in Eq. (\ref{eq:alpha}). The $2k_{F}$ filtered data could best be
fit by the analytical form
\begin{equation}
C2(r)=C\cos (2k_{F}\,r+\varphi _{c})r^{-2\,K_{c}}\exp
(-r/d_{c})\text{ ,} \label{eq:C2a}
\end{equation}
which has a coherence length $d_{c}$ that is related to the
excitation gap in a finite chain. The resulting power law exponent
and coherence length are also shown in Table \ref{table1}. $K_{c}$
is within 5 \% of the calculated values of $K$ obtained by
inverting Eq. (\ref{eq:alpha}). $d_{c}$ seems closely correlated
to $\xi $ and both to $E_{x}^{-1}$. The Pearson correlation
coefficient between $d_{c}$ and $\xi $ is 0.995 and 0.984 between
$d_{c}$ and $E_{x}^{-1}$. The two coherence lengths $\xi $ and
$d_{c}$ enter the fitting functions of Eqs. (\ref{eq:C1b}) and
(\ref{eq:C2a}) in quite different ways. This probably stems from
the different role the block states $\left| \psi ^{B}\right\rangle
$ play in matrix element storage, off diagonal with respect to
total block occupation $\mathfrak{N}^{B}$ in single-fermion
functions $\left\langle \psi ^{B}(\mathfrak{N}_{1}^{B})\left|
c_{j}\right| \psi ^{B}(\mathfrak{N}_{2}^{B})\right\rangle $ but
diagonal for the charge correlations $\left\langle \psi
^{B}(\mathfrak{N}_{1}^{B})\left| n_{j}\right| \psi
^{B}(\mathfrak{N}_{1}^{B})\right\rangle $. One last comment
concerns the $q=0$ charge correlations. We found no evidence for
this contribution in our data possibly because of the specific
quantity we chose to calculate in Eq. \ref{eq:C2}.

\begin{table}
\caption{\label{table1}Various parameters calculated for a
quarter-filled chain at different values of $V$ and for
$N_{B}=42.$.}
\begin{ruledtabular}
\begin{tabular}{ccccccccc}
$V$ & $E_{x}$ & $k_{F}$ & ${\bf \alpha }$ & $K$ & ${\bf \xi }$ & $\Lambda $ %
& $K_{c}$ & $d_{c}$ \\ \hline
1 & 0.049 & 0.78 & 0.106 & 0.63 & 81 & 1.5 & 0.67 & 100 \\
2 & 0.064 & 0.78 & 0.27 & 0.49 & 76 & 1.2 & 0.52 & 87 \\
3 & 0.076 & 0.78 & 0.49 & 0.39 & 65 & 1.2 & 0.42 & 69 \\
4 & 0.088 & 0.78 & 0.76 & 0.31 & 59 & 1.3 & 0.35 & 59 \\
5 & 0.101 & 0.77 & 1.08 & 0.25 & 53 & 1.4 & 0.29 & 53 \\
6 & 0.114 & 0.77 & 1.50 & 0.21 & 47 & 1.5 & 0.20 & 43
\end{tabular}
\end{ruledtabular}
\end{table}

We wish to point out an interesting observation we made on the raw
(unaveraged) occupation $n_{j}$ near the ends. We can fit the
occupation by the relation
\begin{equation}
n_{j}\approx \overline{n}+n_{0}\cos \left( 2k_{F}j+\phi \right)
\left( j\right) ^{-K_{c}}  \label{eq:end_occ}
\end{equation}
where $\overline{n}=0.25$ and $0.25\lesssim n_{0}\lesssim 0.4$.
A similar observation has recently been reported in Ref. \onlinecite{White01}.
The broken symmetry state resulting from the pinning at the chain
ends forces the local occupancy variation $\left\langle \delta
n_{j}\right\rangle $, where $\delta n_{j}=n_{j}-\overline{n}$, to
be equal to the root mean square fluctuation $\sqrt{\left\langle
\delta n_{0}\delta n_{j}\right\rangle }$.

We wish to end this subsection by examining the situation for
$N_{B}=10$. Why such a small number of block states? We have
already stated that our calculations were made with $N_{B}\leq
128$. This is at the limit of our computational capabilities. If
the chains were independent, this would be equivalent to having
some 10 block states per chain, which is not large indeed. At such
small values of $N_{B}$, we had to introduce another coherence
length $\xi _{c}$ for the charge correlations,
\begin{eqnarray}
C2(r) &=&C\cos (2k_{F}\,r+\varphi _{c})  \nonumber \\
&&\times \left( \xi _{c}\sinh (r/\xi _{c})\right) ^{-2\,K_{c}}\exp (-r/d_{c})%
\text{ .}  \label{eq:C2b}
\end{eqnarray}
We used a fitting procedure which weighed more heavily the data
for $r\lesssim\xi $ so as to be able to recover key\ parameters
with values close to those at $N_{B}=42$. Truncation errors run
typically at the level of a few times $10^{-5}$. This is
considerably larger than for $N_{B}=42$. Table \ref{table2} gives
some of the parameters of the fit. We have used the values of
$\Lambda $ of Table I. $\alpha $ and $K_{c}$ compare favorably.
The ground state excitation energy has appreciably increased and
the coherence length has shortened. They are rather featureless, a
signature of the small number of block states. Note that $\xi
_{c}\sim d_{c}$ is somewhat ``elastic'' in the sense that its
value can drift significantly without marked effect (within the 5
\% error bar) on the fit.

\begin{table}
\caption{\label{table2}Various parameters calculated for a
quarter-filled chain at different values of $V$ and for
$N_{B}=10.$.}
\begin{ruledtabular}
\begin{tabular}{ccccccc}
$V$ & $E_{x}$ & ${\bf \alpha }$ & ${\bf \xi }$ & $K_{c}$ & ${\bf \xi_{c} }$ %
& $d_{c}$ \\
\hline
1 & 0.20 & 0.10 & 19 & 0.65 & 30 & 22 \\
2 & 0.23 & 0.26 & 15 & 0.50 & 31 & 19 \\
3 & 0.24 & 0.48 & 16 & 0.41 & 35 & 22 \\
4 & 0.26 & 0.75 & 17 & 0.33 & 34 & 21 \\
5 & 0.27 & 1.10 & 17 & 0.26 & 32 & 21 \\
6 & 0.29 & 1.50 & 15 & 0.21 & 32 & 21
\end{tabular}
\end{ruledtabular}
\end{table}

\section{Coupled Chains}

\begin{figure}
\includegraphics[trim=.55in 2.1in 0.7in 2.5in, scale=.45,clip]{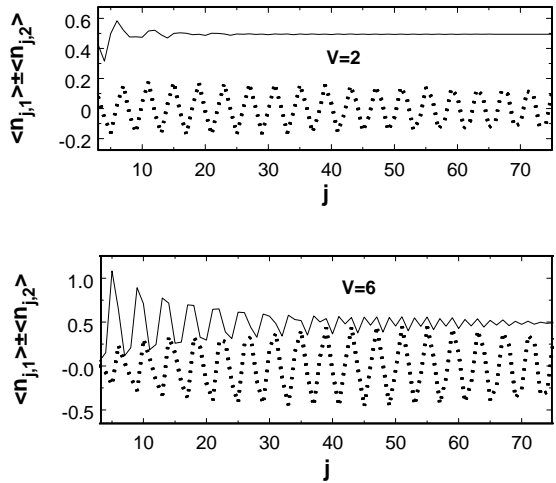}
\caption{\label{fig4}On-site charge (full line, + sign) and
polarization (dotted line, - sign) in typical broken symmetry
states at $V=2$, $6$ and for $t_{\perp }=0.5$.}
\end{figure}

Now that we have acquired sufficient experience and confidence in
data management, we can tackle the study of two coupled chains.
One final observation is warranted. We found that convergence of
our DMRG algorithm
could only be achieved relatively quickly for an odd number of fermions $%
N_{f}=(2N_{f0}+1)$. This can be understood in view of our
discussion in Section \ref{sec:clbf}. First of all, fermions are
pushed to the ends by the Coulomb repulsion. The fermion
occupancies on both chains thus start site-synchronized at the
ends. Secondly, the transverse hopping favors out of phase
occupancies on the chains. Fig. \ref{fig4} illustrates this. It
shows the on-site charge $\left( \left\langle n_{j,1}\right\rangle
+\left\langle n_{j,2}\right\rangle \right) $ and polarization
$\left( \left\langle n_{j,1}\right\rangle -\left\langle
n_{j,2}\right\rangle \right) $ that is typical of the
broken-symmetry ground state. The larger $V$ is, the more
pronounced the out of phase character (polarization) is and the
larger the $4k_{F}$ charge component. This behavior can be
achieved more easily in a state in which one chain has one extra fermion. The
chain with $(N_{f0}+1)$ fermions is more compressed and cannot easily sustain
solitons. But then, the chain with $N_{f0}$ fermions can easily
accommodate a kink-antikink pair which then allows the inner part
of this chain to be out of phase with the ends and with the other
chain. Broken symmetry states also have the advantage of focusing
the computational resources to a single non-degenerate state
instead of splitting them between degenerate states thus
decreasing the numerical coherence length. We have thus chosen to
do our calculations for $N=150$. Our truncation error varies from
$5\times 10^{-5}$ for $N_{B}=42$ to $5\times 10^{-6}$ for the
larger values $N_{B}=128 $. We have kept the $\Lambda $ values of
the single chain.

We shall first look at single-fermion behavior and then at some
two-fermion correlation functions.

\subsection{Single-fermion transfer function}

The perturbative renormalization group formulation in Ref.
\onlinecite{Bourbon91} presented a unified description of the
renormalization of $t_{\perp}$ and the generation of interchain
couplings in quasi-1D solids. Although the basic elements are
present, the treatment is perturbative and subject to caution for
large interactions. The bosonization approaches have the potential
to do better in this respect since the single-chain interacting
spinless fermion problem has an exact solution.

Let us then examine the predictions of bosonization. There are two
key treatments which look at our Hamiltonian from two different
perspectives. Nersesyan {\it et al.}\cite{Nersesyan93} (NLK)
bosonize the chain fermion operators $c_{j,\beta }$, whereas
Yoshioka and Suzumura\cite{Suzumura95} (YS) do so with the band
operators
\begin{equation}
c_{j,\sigma }=\left( c_{j,1}+\sigma c_{j,2}\right) /\sqrt{2}\text{
}. \label{eq:bandop}
\end{equation}
Here $\sigma =\pm 1$ is the band index. One has $k_{\perp
}=(1-\sigma )(\pi /2).$ The procedure yields two separated
sectors, polarization and occupation, that, by analogy to a chain
of spins  $\frac{1}{2}$ in a magnetic field $\varpropto t_{\perp
}$, are labelled spin and charge.

In NLK, there is spin-charge separation and the Coulomb
interaction is absorbed within the stiffnesses $K_{c}$ and
$K_{s}$. Here $K_{c}$ has the same value as for a single chain. $\
t_{\perp }$ appears in the spin sector and acts as the generator
of interchain two-fermion couplings $G$ and
$\widetilde{G}$ corresponding to particle-hole and particle-particle pair
hopping. The renormalization group (RG) equations for $G$ reads $%
G^{\prime }=2(1-K_{s})G+(K_{s}-\widetilde{K}_{s})\tau ^{2}$ ,where
$\tau =\left| t_{\perp }\right| \,\Lambda /(2\pi u_{s})$, $u_{s}$
is the spin excitation velocity, $\widetilde{K}_{s}=1/K_{s}$, and
the prime indicates the derivative with respect to $\ell =\ln
\left( \max \left( \omega /E_{0},T/E_{0},v_{F}k/E_{0}\right)
\right) $ where $E_{0}$ is the starting
energy scale. This is discussed in Refs. \onlinecite{Solyom79} and %
\onlinecite{Bourbon91}. The equation for $\widetilde{G}$ is
obtained by the substitution $\widetilde{K}_{s}\rightleftarrows K_{s}$. Note
that $G(\ell =0)=\widetilde{G}(0)=0$. This RG equation is different from
Yakovenko's\cite{Yakovenko92} whose $\tau ^{2}$ term is larger by a factor $%
8\pi ^{2}$. This does not change the qualitative behavior of the
equations, only the numbers. Furthermore, it is more in line with the
coefficients of the RG equations
in Ref. \onlinecite{Bourbon91}. The RG also renormalizes $t_{\perp
}$, $\tau ^{\prime }=(2-\Delta _{s})\tau $ where $\Delta _{s}=
\frac{1}{2}(K_{s}+\widetilde{K}_{s})$, and $K_{s}$ is governed by $(\ln
K_{s})^{\prime }=\frac{1}{2}\left(
\widetilde{K}_{s}^{2}\widetilde{G}^{2}-K_{s}G^{2}\right) $. Note
that one has $K_{s}(\ell =0)=K_{c}$. \ There are additional
contributions of order $\tau ^{2}$ to this last equation\cite{Bourbon91}, a
fact acknowledged by NLK, but which will remain unexplored by us. When $G$
reaches strong coupling ($G \sim 1$) then a gap opens in the spin sector and
only the charge sector contributes to power laws.

In YS, there is also spin-charge separation for small $t_{\perp
}$. But the authors point out that this is no longer true for
large transverse hopping, when the Fermi velocities for the two
bands are appreciably different. This should be kept in mind as
our values of $t_{\perp }\geq 0.1$ should qualify. The charge
sector behaves essentially as in NLK, with $\eta \equiv K_{c}$.
The spin sector, however, transforms the Coulomb interaction into
two inter-band couplings $g_{2\phi +}$ and $g_{2\phi -}$ such that
the spin stiffness $\eta _{\phi }=1$ at $\ell =0$. This
dichotomy between spin and charge sectors is satisfactory for
small $V$. The RG equations (when corrected for typographical
errors) yield solutions qualitatively similar to NLK for $t_{\perp
}(\ell )$. But it rapidly becomes annoying at large interactions
since the spin sector quickly, too quickly, goes to strong
coupling when $g_{2\phi +}$ or $g_{2\phi -}\sim 1$ and develops a
gap. It then appears that the NLK approach should do better before
the single-fermion dimensionality crossover and YS after the
crossover, when bands have formed.

\begin{figure}
\includegraphics[trim=0in 2.2in 0in 2.9in, scale=.4,clip]{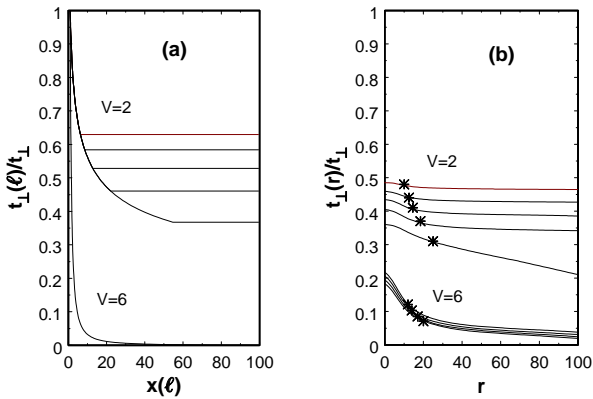}
\caption{\label{fig5}Transverse hopping as a function of the
length scale normalized to the bare value for various situations.
The curves for $V=2$ are for $0.5\geq
t_{\perp }\geq 0.1$ from top to bottom. (a) NLK bosonization results. The $%
V=6$ curves are for all values of $t_{\perp }$. (b) DMRG results.
The $V=6$ curves are for $0.5\geq t_{\perp }\geq 0.2$, from top to
bottom,
respectively. The stars are set at the values of the coherence length $%
\protect\xi .$}
\end{figure}

Let us now focus on the predictions of these models for the
transverse hopping as a function of length scale $x(\ell )=\exp
(\ell $ )$.$ We first start with NLK at $\ell =0$. The RG
equations will switch to YS once the dimensionality crossover is
reached, when $\tau (\ell _{x})\sim (2\pi )^{-1}$, via the mapping
proposed in NLK. We have chosen this value as it produces
crossovers at scales comparable with the ones of Bourbonnais\cite
{Bourbon91} and YS. With this choice, the renormalized transverse
hopping
\[
t_{\perp }(\ell )=(2\pi u_{s}/\Lambda )\tau (\ell )\exp (-\ell
)\approx 2\pi v_{F}\tau (\ell )/x(\ell )
\]
leads to a dimensionality crossover at
\begin{eqnarray}
t_{\perp }(\ell _{x}) &=&(2\pi u_{s}/\Lambda )\tau (\ell _{x})\exp
(-\ell
_{x})  \nonumber \\
&\approx &v_{F}\exp (-\ell _{x})=v_{F}/x(\ell _{x})\text{ ,}
\label{eq:crossover}
\end{eqnarray}
in which we have set $\Lambda =1$ and $u_{s}=v_{F}$ . This is the
result of Bourbonnais for the crossover. Fig. \ref{fig5}(a) shows
the results for the
renormalized transverse hopping $t_{\perp }(\ell )$ at both intermediate ($%
V=2$) and strong ($V=6$) Coulomb interactions. This was obtained using the $%
K_{s}=K$ values in Table \ref{table1} at $\ell =0$. On the one
hand, the $V=2 $ results are typical of small $V$ and large
$t_{\perp }$ and clearly show the dimensionality crossover. The
``ladder'' diagram is typical of this crossover regime. The
renormalized hopping after the crossover satisfies the relation
\begin{equation}
\left( t_{\perp }(\ell _{x})/t_{\perp }\right) \varpropto \left(
t_{\perp }\right) ^{\alpha /(1-\alpha )}  \label{eq:renormlaw}
\end{equation}
as proposed by Bourbonnais and verified by Capponi. On the other hand, the $%
V=6$ curves for different $t_{\perp }$ all superimpose, a
signature of the confined regime ($\alpha>1$).

The DMRG calculations we now present were taken at $N_{B}=96$. The
following band transfer functions were calculated
\begin{equation}
C1(r,\sigma )=\overline{C1((N-r)/2,r,\sigma )}  \label{eq:C1c}
\end{equation}
where $C1(j,r,\sigma )=\langle c_{j+r,\sigma }^{\dagger
}c_{j,\sigma }\rangle $. Each of these two $C1(r,\sigma )$ was
fitted with Eq. (\ref {eq:C1b}). The fit could only be made by
allowing the $k_{F}(r,\sigma)$ which characterize the bands $\sigma$
to vary as a function of $r$, a situation not
seen in single chains. This is how the renormalization of the
transverse hopping manifests itself most directly. We could then
calculate the renormalized transverse hopping from
\begin{equation}
t_{\perp }(r)\approx \left[ k_{F}(r,\sigma =1)-k_{F}(r,\sigma
=-1)\right] /v_{F}\text{ .}  \label{eq:tperp}
\end{equation}
The results for this last quantity are reproduced in Fig.
\ref{fig5}(b). The four upper curves for $V=2$ clearly show the
``ladder'' characteristic of the crossover situation. This
crossover, from Eq. (\ref{eq:crossover}), is approximately at
$r_{x}\approx $ $v_{F}/t_{\perp }(r_{x})$ which has been reached
before the coherence length is reached for $t_{\perp }\geq 0.2$.
The renormalized transverse hopping is also rather flat for
$r>r_{x}$. This is not the case for the $t_{\perp }=0.1$ curve
which has a ``drooping'' (negative slope) characteristic. This
intermediate value of $V$ has this nice property of showing both
the crossover and its absence due to numerical coherence.
Moreover, the predicted renormalization law of Eq. (\ref
{eq:renormlaw}) gives a value $\alpha \sim 0.26$ for the top four
curves, a value in excellent agreement with the single chain one.
This is
somewhat accidental as the same analysis performed on the $V=1$ data for $%
0.2\leq t_{\perp }\leq 0.5$ yields a value of $\alpha \sim 0.14$,
somewhat larger that its value in a single chain. The problem, if
there is one, may lie with Eq. (\ref{eq:tperp}) which does not
account for the change in Fermi velocity that occurs when the
values of $k_{\perp }=0,\pi $ are appreciably different (large
$t_{\perp }$) or due to renormalization. One must bear in mind
that the DMRG is on a lattice and the fermion energy is far from
the linear dispersion of the RG. Non-universal band edge effects
(pre-RG) might be considerable and this might explain the
discrepancy at the sizeable values of $t_{\perp }$ we have used.
We find no clear evidence for a crossover at $V\geq 3$ as $\xi
<r_{x}$ and the $t_{\perp }(r)$ curves do not flatten out at
$r_{x}$.

The $V=6$ curves are bunched together and show confinement. This
is in qualitative agreement with bosonization. Our results
indicate that the coherence length sets the scale for the
evolution of the transfer function.
Further evolution, such as crossovers, are blocked at distances beyond $\xi $%
. The splitting, albeit small, of the different $V=6$ curves is caused by just
this effect. The coherence lengths are shorter for the larger
values of $t_{\bot }$ thus leading to larger, that is less
renormalized, values of $t_{\perp }(\ell )$.

It is important to understand the connection between $t_{\perp }(r)$ and $%
t_{\perp }(\ell )$. $t_{\perp }(r)$ can be viewed, from Ref. %
\onlinecite{Bourbon91}, as an average over all momentum scales
$k\lesssim
k_{r}\equiv 2\pi /r$. It is thus an average of $t_{\perp }(\ell )$ over $%
\ell \lesssim \ln (v_{F}k_{r}/E_{0})$. Thus clearly $t_{\perp }(r=0)=\overline{%
t_{\perp }(\ell )}\neq t_{\perp }(\ell =0)$. That is why $t_{\perp
}(r=0)$ is larger than $t_{\perp }(r)$ yet smaller than the bare
value $t_{\perp }$.

\begin{figure}
\includegraphics[trim=0in 2.2in 0in 2.9in, scale=.4,clip]{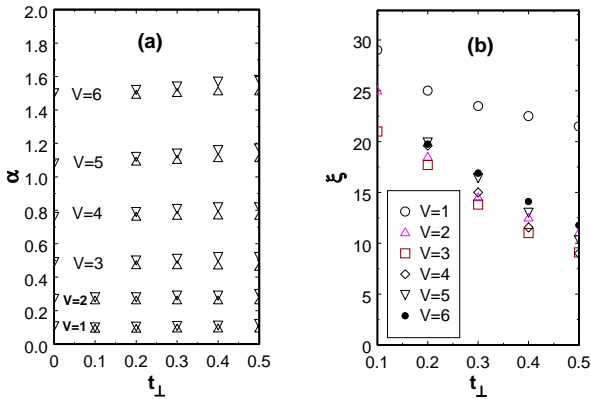}
\caption{\label{fig6}Fitted values for the DMRG transfer function
at different transverse hopping and Coulomb interaction strengths.
(a) Power law exponent results. The downward pointing triangles
are for $k_{\perp }=\protect\pi $ while the upward pointing ones
are for $k_{\perp }=0$. (b) Mean values of the numerical coherence
length.}
\end{figure}

Fig. \ref{fig6}(a) shows the fitted values for $\alpha $ for both
$k_{\perp }=0,\pi $. They are surprisingly nearly independent of
$t_{\perp }$ or of the existence or not of a dimensionality
crossover. This is not expected from
bosonization which would predict a variation of $K_{s}(\ell )$ and thus of $%
\alpha $. This is perhaps a consequence of the absence of true
spin-charge separation. YS also predict a change in exponent at
the crossover. Fig. \ref {fig6}(b) shows the average numerical
coherence lengths for $k_{\perp
}=0,\pi $. The coherence lengths monotonically decrease at constant $V$ as $%
t_{\perp }$ increases. Curiously, $\xi $ is smallest at $V=2,3$.
We will come back to this in the next subsection.

\begin{figure}
\includegraphics[trim=0in 2.4in 0in 3.1in, scale=.4,clip]{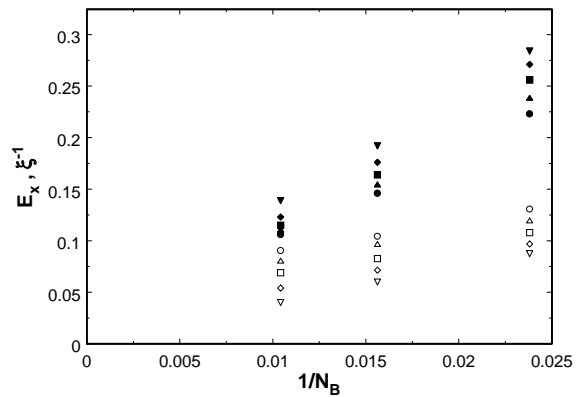}
\caption{\label{fig7}Ground state excitation energy $E_{x}$ (full
symbols) and inverse
coherence length $\protect\xi ^{-1}$ (empty symbols) as a function of $%
N_{B}^{-1}$ for $V=2$. The curves are for $0.5\geq t_{\perp }\geq
0.1$ from bottom to top, for $E_{x}$, and top to bottom for
$\protect\xi ^{-1}.$}
\end{figure}

\begin{figure}
\includegraphics[trim=0in 2.4in 0in 3.1in, scale=.4,clip]{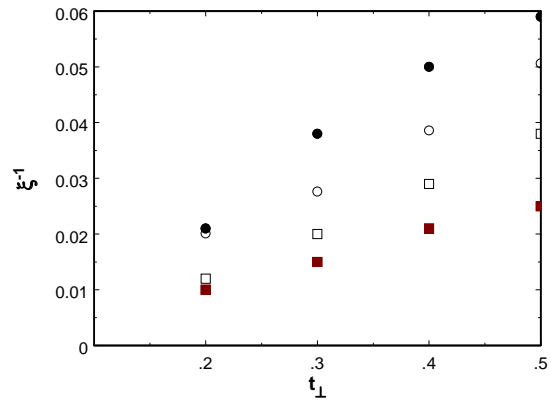}
\caption{\label{fig8}Inverse coherence length extrapolated to an
infinite number of block states, circles, and reciprocal of the
thresholds to strong coupling calculated from the bosonization
equations, squares, as a function of the transverse hopping. The
results for $V=2$, full symbols, and $V=6$, empty symbols, are
shown.}
\end{figure}

Quite obviously, it would be futile to study the scaling behavior
of these quantities with the chain length since $\xi $ is much
shorter. But it would be possible to study the scaling with
$\left( N_{B}\right) ^{-1}$ which acts like a temperature. In an
attempt to better understand our results, we thus repeated the
calculations at other values of $N_{B}$. Fig. \ref{fig7} shows the
scaling behavior of the ground state excitation energy and the
reciprocal of the coherence length for $V=2$. The gap is seen to
extrapolate, as $N_{B}\rightarrow \infty $, to small values of
$E_{x}$
compatible with $N=150$. There is thus no indication of an energy gap. But $%
\xi ^{-1}$ no longer correlates well with $E_{x}$ in contrast to
results for single chains. As a matter of fact, the extrapolated
value increases
quasi-linearly with $t_{\perp }$. The same behavior is found for all values $%
V\geq 1$. Fig. \ref{fig8} shows the $N_{B}\rightarrow \infty $
extrapolated inverse coherence lengths for $V=2$ and $V=6$ as well
as the reciprocal of the threshold values to strong coupling
coming from numerical solutions of the bosonization equations. We
see there is a tight correlation between bosonization and DMRG
behaviors with $t_{\perp }$, though the length scales are
different, in both situations with or without a crossover. This
suggests the coherence lengths are affected by the oncoming strong
coupling regime at which our fitting formula, Eq. (\ref{eq:C1b}),
would no longer be valid.


\subsection{Two-particle correlations}

Following Eq. \ref{eq:C2}, we define the on-site charge ($\nu =+$)
and
polarization ($\nu =-$) correlation functions $C2(r,\nu )=\overline{%
C2((N-r)/2,r,\nu )}$, where
\begin{equation}
C2(j,r,\nu )=\left\langle \left( n_{\nu ,j+r}-\left\langle n_{\nu
,j+r}\right\rangle \right) \left( n_{\nu ,j}-\left\langle n_{\nu
,j}\right\rangle \right) \right\rangle \text{ ,}  \label{eq:corr}
\end{equation}
and
\begin{equation}
n_{\pm ,j}=n_{j,1}\pm \,n_{j,2}\text{ .}  \label{eq:operator}
\end{equation}
What should we expect for the power law exponents
\begin{equation}
C2(r,\nu )\varpropto r^{-K_{\nu }}  \label{eq:correxp}
\end{equation}
characterizing these correlation functions? It is quite clear that as $%
t_{\perp }\rightarrow 0$, the chains become independent and one should have $%
K_{\nu }=K_{c}+K_{s}=2K_{c}$. This is consistent with NLK. From
the band
perspective, the approach of YK predicts $K_{+}=(\eta +\eta _{\phi })$ and $%
K_{-}=(\eta +1/\eta _{\phi })$ when there is no gap. Since $\eta
_{\phi }>1$ for repulsive interactions, one should expect
$K_{+}>K_{-}$.

\begin{figure}
\includegraphics[trim=0in 2.4in 0in 2.9in, scale=.4,clip]{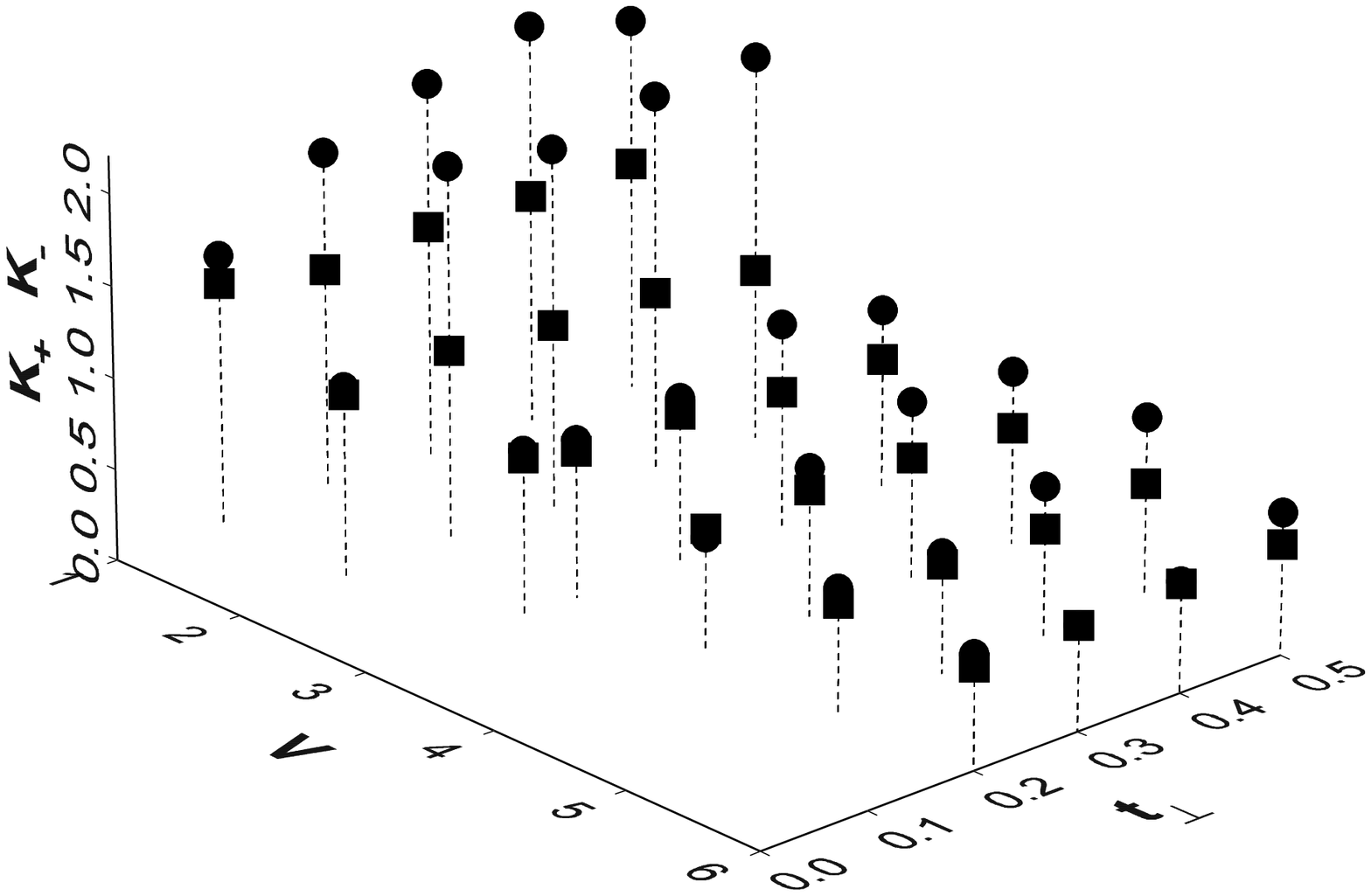}
\caption{\label{fig9}Power law exponents for charge, full circles,
and polarization, full boxes, for various values of $V$ and
$t_{\perp }$.}
\end{figure}

We have calculated these correlation functions and tried fitting
them to forms reminiscent of Eq. (\ref{eq:C2b}). The polarization
correlation function behaves nicely and can be fitted with
\begin{eqnarray}
C2(r,-) &=&C_{-}\cos ((k_{F}(r,1)+k_{F}(r,-1))r+\varphi _{-})  \nonumber \\
&&\times \left( \xi _{-}\sinh (r/\xi _{-})\right) ^{-K_{-}}\exp
(-r/d_{-}) \label{eq:polarfit}
\end{eqnarray}
using the filtering procedure explained in the data processing of
single-chain correlations. The exponents $K_{-}$ are essentially
those of the charge correlations of single chains as witnessed in
Fig. \ref{fig9}. Again, this is unexpected. It is as if $(\eta
+1/\eta _{\phi })$ or $\left( K_{c}+K_{s}\right) $ remained nearly
constant. This behavior is like the one observed for the single
particle exponent $\alpha $. This points to the breakdown of
spin-charge separation for a highly non-linear dispersion. As for
$\xi _{-}$ and $d_{-}$, we find they are generally larger than
$\xi $, sometimes by a sizeable factor, and also fairly
``elastic'' as observed for single chains.

\begin{figure}
\includegraphics[trim=0in 2.4in 0in 3.1in, scale=.4,clip]{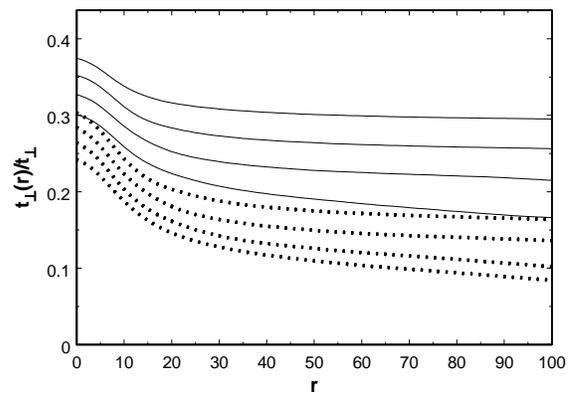}
\caption{\label{fig10}Transverse hopping as a function of the
length scale normalized to the bare value for $V=3$ (full lines)
and $V=4$ (dotted lines). The curves are for $0.5\geq t_{\perp
}\geq 0.2$ from top to bottom.}
\end{figure}

The charge correlation function proved a bit more subtle to fit.
The charge correlations generally decrease much faster than the
polarization correlations. They rapidly become quite noisy when
they reach the limits set by the truncation error. This can easily
be monitored on a logarithmic plot of the amplitude of $C2(r,+)$
as a function of $r$. There is a sudden break in the general
linearlike decrease. The range of values in which a function
having an amplitude of the type proposed in Eq.
(\ref{eq:polarfit}) can be appropriate is often limited to
$r\lesssim 30$ when the coherence lengths are small. We have
extended our calculations to $N_{B}=128$ for a few cases as a
check of the sturdiness of our calculations. It is mostly $\xi
_{-}$ that is affected by a noticeable increase. Curve fitting
with many parameters can thus become delicate and leads to large
uncertainties which we estimate at 10\% for $K_{+}$. We found that
we could fit $C2(r,+)$ with a form equivalent to Eq.
(\ref{eq:polarfit}) with $\nu =+$ instead of $\nu =-$
for all cases $V\geq 3$. The fast Fourier transform reveals only one broad ``%
$2k_{F}$'' wavenumber. $K_{+}$ is always of the same size or a bit
larger than $K_{-}$ in this chain-like regime as can be seen in
Fig. \ref{fig9}. For those situations $V\leq 2$, the fast Fourier
analysis reveals two wavenumbers associated with $2k_{F}(r,1)$ and
$2k_{F}(r,-1).$ We sometimes fitted the most prominent wavenumber
or sometimes both, when the separation was not big enough, with
a form that combines contributions of both bands (see Eq. \ref{eq:bandop})
\begin{eqnarray} C2(r,+) &=&\sum_{\sigma }C_{+,\sigma
}\cos (2k_{F}(r,\sigma )\,r+\varphi
_{+,\sigma })  \nonumber \\
&&\times \left( \xi _{+}\sinh (r/\xi _{+})\right) ^{-K_{+}}\exp (-r/d_{+})%
\text{ .}  \label{eq:chargefit}
\end{eqnarray}
As can bee seen from Fig. \ref{fig9}, $K_{+}$ is appreciably larger than $%
K_{-}$ in the low $V$ regime. We found $\xi _{+}$ and $d_{+}$ to
be somewhat smaller than $\xi $ and still ``elastic''. It is quite
apparent from Fig.
\ref{fig9} that there is a discontinuity in $K_{+}$ going from $V=2$ to $V=3$%
, with values dropping from 2 to 1. This behavior is, we believe,
intrinsic to the coupled chains and is not a result of short
coherence lengths since these are of the same size for $V=2,3$. It
indicates a crossover from a single-fermion dominated behavior,
when the charge fluctuations are propagated through independent
band-like particle and hole motion showing two wavenumbers in the
charge response function, to a pair behavior, when the
fluctuations are carried by particle-hole pairs having chain-like
character sharing a common wavenumber. Such a crossover was
deduced by Capponi\cite{Capponi98} in the same range of values of
$V$. In this crossover region, there is fierce competition between
single-fermion behavior and coherent pair motion. The computer
resources are shared between the two competing behaviors and this
explains the shorter coherence lengths in the crossover region. In
spite of the domination of the pair motion, the transverse hopping
is not necessarily confined as Fig. \ref{fig10} shows. There are
no clear crossovers at the larger $t_{\perp }$, using the criteria
discussed in the previous subsection, but the hopping is
considerable and decreases only slowly with distance. There may
well be coexistence between coherent interchain pairing and some
sort of interband single particle behavior before the
strong-coupling limit is reached, a possibility raised by NLK.

It should finally be mentioned that, as was described in Section
\ref{sec:dataproc} for single chains, we find a
broken symmetry state in which the total on-site occupancy variation $%
\left\langle \delta n_{+,j}\right\rangle$ is proportional to the
square root of the charge correlation amplitude $\left( \xi
_{+}\sinh (j/\xi _{+})\right) ^{-K_{+}/2}\exp (-j/2d_{+})$ near
the ends.

\section{Summary}

Let us now sum up the more important findings on the
quarter-filled double-chain problem. Our working hypothesis is
that the fitting functions and the data treatment we have used for
the analysis of one- and two-particle functions, which were
validated for single chains, can safely be carried over to two
coupled chains.

1. The central result, which motivated this whole study, is the
confirmation of strong renormalization of the inter-chain hopping
in the presence of the Coulomb interaction $V$. This was measured
by monitoring the difference in the Fermi momentum of the two
bands beyond the dimensionality crossover. The extent of the
renormalization is much stronger that expected from the RG or
bosonization at small $V$. It is considerable at the larger $V$.
It is most likely going to zero as can be concluded by observing
the trend in $t_{\perp }(r)/t_{\perp }$ which decreases with
$t_{\perp }$. The proposed crossover law, Eq.
(\ref{eq:renormlaw}), was seen to be only approximately valid when
the crossover exists. It is probably our definition, Eq.
(\ref{eq:tperp}), and band edge effects that are responsible.

2. The single-fermion transfer function power-law exponent $\alpha
$ and the polarization correlation function exponent $K_{-}$ were
seen to be essentially those of the single chain. This is
interpreted as a sign of possible violation of spin-charge
separation.

3. The charge correlation function shows superposition of two
wavenumbers and has an exponent $K_{+}$ that is twice as large as
the one for polarization for $V\leq 2$. This indicates the
existence of a crossover from independent particle-hole motion to
correlated pair motion between $V=2$ and $V=3$.

4. The ground state excitation energy, when extrapolated to $%
N_{B}\rightarrow \infty $, is as expected for a finite system of
150 sites. The coupled chains thus remain gapless. There are also
indications, in the same limit, of possible strong coupling
regimes as inferred from the linear-like $t_{\perp }$ dependence
of the extrapolated inverse coherence lengths, much in the same
way that bosonization predicts.

The rather short numerical coherence lengths we have encountered
with the DMRG, although expected from the study on single-chains,
did put some stress on the curve fitting procedures. This resulted
in fairly large errors in our exponents.



\end{document}